\documentclass[11pt]{article}

\usepackage{endnotes}            
\usepackage{amsfonts}            
\usepackage{amssymb}             
\usepackage{amsthm}              
\usepackage{amsmath}            
\usepackage{algorithm}           
\usepackage{algorithmic}
\usepackage[letterpaper]{geometry}
\usepackage{rotating}            
\usepackage{graphicx}
\usepackage{xcolor}
\usepackage{multibib}

\usepackage{setspace}                
\usepackage{natbib}
\bibpunct{(}{)}{;}{a}{,}{;}

\begin{document}

\title{
The Decay of Impact with Network Distance in Linear Diffusion Processes\thanks{This work was supported under NSF award SES-2448652.}
}

\author{
Alexander Murray-Watters\thanks{Department of Sociology, University of California, Irvine} \and 
Cheng Wang\thanks{Department of Sociology, Wayne State University}
\and
John R. Hipp\thanks{Department of Criminology, Law, and Society, University of California, Irvine}
\and
Cynthia Lakon\thanks{Department of Health, Society, and Behavior Administration, University of California, Irvine}
\and
Carter T. Butts\thanks{Departments of Sociology, Statistics, Computer Science, and EECS and Institute for Mathematical Behavioral Sciences, University of California, Irvine}\thanks{To whom correspondence should be addressed. SSPA 2145; University of California, Irvine; Irvine, CA 92697; \texttt{buttsc@uci.edu}}
}
\date{4/5/26}
\maketitle

\begin{abstract}
  Many processes related to status, power, and influence within social networks have been modeled using forced linear diffusion models; examples include the highly successful Friedkin-Johnsen model of social influence, the status/power scores of Katz and Bonacich, and the widely used network autocorrelation model.  While a basic assumption of such models is that the impact of one individual on another through any given path falls exponentially with path length, the total impact of the first individual on the second involves contributions from walks of all lengths; thus, while total impact is expected to decline with network distance, the relationship is not trivial.  Here, we provide an approximate solution for the total impact of one node on another as a function of network distance, showing that the total impact is given to first order by a product of eigenvector centrality scores together with an expression in terms of the graph spectrum (eigenvalues of the adjacency matrix) that falls exponentially with distance.  We also show how this solution can be refined using higher-order eigenvectors of the adjacency matrix.  A numerical study on interpersonal networks drawn from educational settings verifies an average exponential decline in impact strength under the linear diffusion model, and shows that the first-order eigenvector approximation can often be a good proxy for total impact as obtained from the exact solution.  This suggests a simple model that can be used to approximate total impact for social influence or status processes in a range of settings.\\[5pt]
\end{abstract}

\theoremstyle{plain}                        
\newtheorem{axiom}{Axiom}
\newtheorem{lemma}{Lemma}
\newtheorem{theorem}{Theorem}
\newtheorem{corollary}{Corollary}

\theoremstyle{definition}                 
\newtheorem{definition}{Definition}
\newtheorem{hypothesis}{Hypothesis}
\newtheorem{conjecture}{Conjecture}
\newtheorem{example}{Example}

\theoremstyle{remark}                    
\newtheorem{remark}{Remark}


Perhaps the most successful model in the social network field is the \emph{forced linear diffusion model} (FLDM), a simple, discrete-time model in which individual states are updated by a weighted average of their neighbors' states, together with an exogenous forcing term reflecting external sources of influence (Eq.~\ref{eq_lindiff}).  While not every network researcher may be familiar with its formal expression, nearly all workers in the field have likely encountered it in one form or another.  The FLDM appears in many guises, and has been used to model a wide range of phenomena (including social influence \citep{friedkin.johnsen:jms:1990}, status and prestige \citep{katz:p:1953,bonacich:jms:1972}, and power \citep{bonacich:ajs:1987,salancik:asq:1986}).  The FLDM is the often-implicit basis of the feedback centrality measures \citep{koschutzki.et.al:ch:2005}, the network (aka spatial) autoregression model \citep{cliff.ord:bk:1973,anselin:bk:1988,doreian:ch:1989}, and the famous PageRank algorithm \citep{brin.page:cnIs:1998}, and explicitly serves as the backbone of the highly successful Friedkin-Johnsen model of social influence \citep{friedkin.johnsen:jms:1990,friedkin:bk:1998,friedkin.et.al:s:2016}.  The FLDM owes its wide utility both to its role as a substantively plausible approximation to many social processes, and to its tractability: many of its properties can be determined or approximated analytically, providing insight and greatly facilitating its use.

\begin{figure}
  \centering
\includegraphics[width=0.6\textwidth]{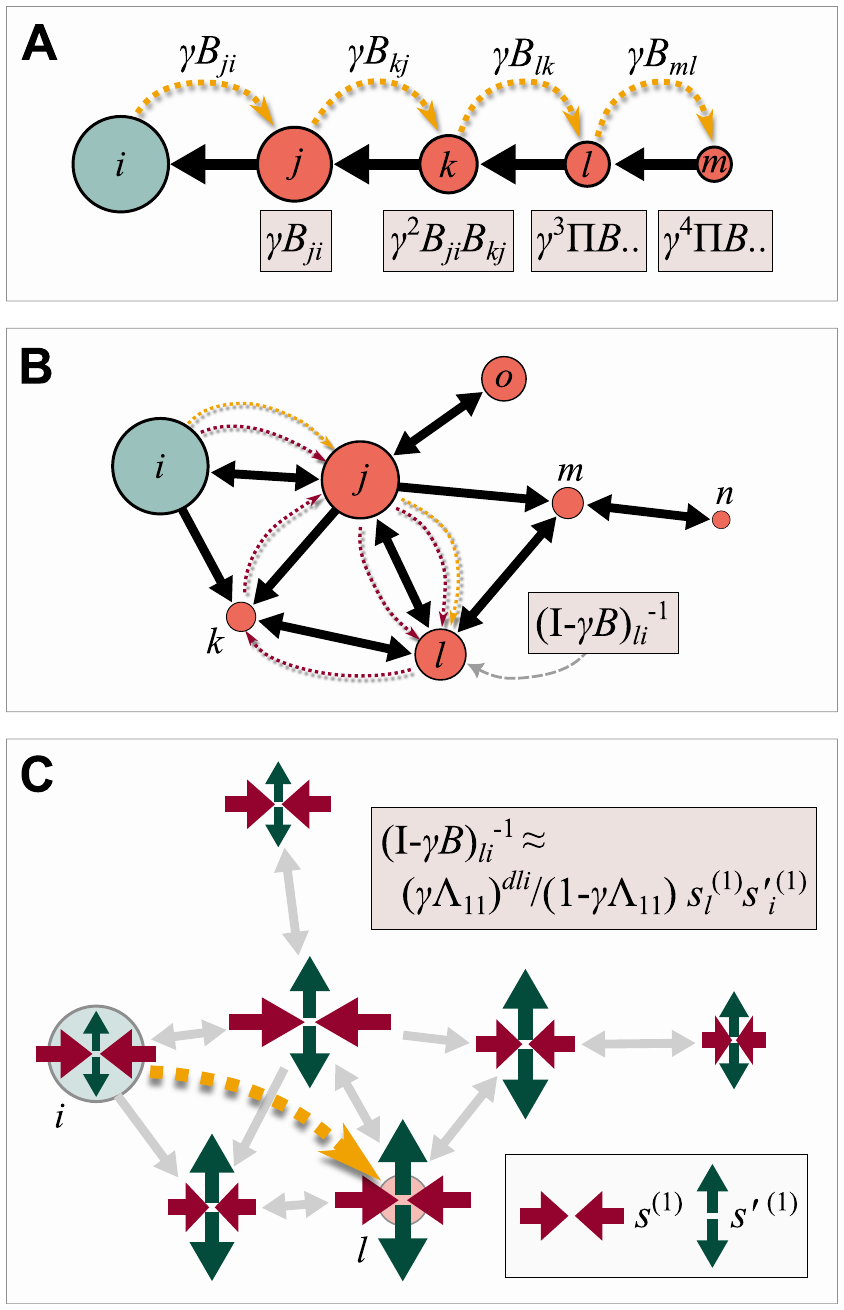}
\caption{Network diffusion under the FLDM.  (A) Impact of node $i$ on alters within a single walk falls exponentially. (Note that, by convention, impact (dotted curves) moves opposite tie direction.)  (B) In general graphs, there may be many walks through which $i$ could affect some node $l$ (dotted lines).  Net impact is a complex function of network structure.  (C) We show that net impact can be approximated by a product of capacities to send (vertical arrows) and receive (horizontal arrows) impact with an exponentially declining function of network distance, with a rate that depends on global structure. (Notation follows Sec.~\ref{sec:model}.)}
\label{fig:fldpworks}
\end{figure}

One such property of substantive interest is the manner in which, in
equilibrium, the total \emph{impact} of ego on an arbitrary alter
declines with distance in the underlying network - i.e., to what
extent can ego ``reach through'' the network to exert a \emph{net} effect on
others?\footnote{Note that our use here refers to the effect of ego on
  alter through the network; this should not be confused with the
  ``impact'' of Social Impact Theory \citep{latane:ap:1981}, which is
  not based on the same underlying model.} This is consequential for e.g.
  anticipating the
potential reach of network interventions, understanding how exogenous
influences on a network may propagate, or mapping the prominent
sources of influence on a given individual.  In recent work, \citet{wang.et.al:s:2026} 
report exponential decay of influence from interventions in a stochastic
actor-oriented model, motivating the question of whether this also
holds for the more tractable FLDM. Superficially, this seems obvious: a simple 
characteristic
of the FLDM is that the direct impact of ego on alter through any given path
(or walk) falls (\emph{ceteris paribus}) exponentially with the path
(or walk) length (see Fig.~\ref{fig:fldpworks}A).  It might thus be
natural to assume that the \emph{total} impact of ego on alter would
likewise decrease exponentially as the distance between them
increases.  However, the total impact at equilibrium of ego on alter
includes not only the contribution of the direct ego-alter geodesic,
but the contributions of all walks of all lengths.  The numbers and
lengths of such walks are not solely a function of the ego-alter
distance, but depend upon the global network structure
(Fig.~\ref{fig:fldpworks}B).  Thus, the total ego-alter impact is a
non-trivial quantity; while it is easily derived, its properties have
not, to our knowledge, been previously characterized in the
literature.

In this paper, we derive an approximation of the total equilibrium impact of one node on another in the FLDM.  This approximation reduces to two elements: a product of eigenvector centralities reflecting the respective net tendencies of the sender and receiver to send and receive impact through the global network structure, and an exponential distance decay whose magnitude is related to the graph spectrum.  This ``field-like'' model (Fig.~\ref{fig:fldpworks}C) thus surprisingly recapitulates the functional form (though not the quantitative aspects) of the direct influence decay, while also incorporating node-specific factors directly related to well-known centrality indices.  In addition to providing conceptual clarity on the primary drivers of impact, this simplified model is potentially amenable to further approximation on incompletely observed networks via e.g. graph sampling, and to calculation on large networks for which the matrix inversions required for exact calculation of the FLDM impact effects are infeasible.  As we show, the approximation can also be improved by adding additional, higher-order terms.

To assess the practical performance of our approximation, we conduct a numerical study using a sample of friendship networks from educational settings.  We find that, on average, total impact does fall approximately exponentially over reasonable network distances, and the first-order approximation to the exact FLDM impact generally performs well.  Adding an additional term further improves the approximation, although the marginal gains over the first-order approximation are relatively modest.

These results shed further light on the local and non-local properties that shape impact in processes ranging from social influence to exchange power, including further elaboration on the role of eigenvector centrality (long known to be deeply tied to status and influence processes).  They also provide a simple, intuitive, and easily calculated  approximation of interpersonal impact that can be straightforwardly applied in any setting for which the FLDM arises.

The remainder of the paper proceeds as follows.  Section~\ref{sec:background} reviews the FLDM, as well as some of the results to be used in our development.  Section~\ref{sec:model} then introduces our approximate total impact model.  The model is then evaluated in Section~\ref{sec:assess} by application to a sample of interpersonal networks; further issues are discussed in Section~\ref{sec:discuss}, and Section~\ref{sec:conclusion} concludes the paper.
  
\section{Background} \label{sec:background}

Many processes of social interest can be approximated by a system in
which: (1) each vertex is associated at any given point of time with a
real-valued state variable ($y$); (2) the state of each vertex is
impacted at each time point by an exogenous \emph{forcing term},
$z$; and (3) the state of each vertex is also influenced at each point
in time by a weighted sum of its neighbors' states within a network,
$G$.  Time is here taken to be discrete, and the evolution of $y$ is
assumed to be time-scale separated from the evolution of $G$ and $z$
(which can thus be treated as fixed).  Taking $A$ to be the adjacency
matrix of $G$, with $W=w(A)$ the matrix of impact weights and $y$
and $z$ to be $N$-vectors (with $N=|V(G)|$ the order of $G$), this
corresponds to the first-order system of difference equations
\begin{equation}
y^{(t)} = W y^{(t-1)} + z, \label{eq_lindiff}
\end{equation}
\noindent noting the directional convention that $W_{ij}$ gives the
immediate impact of $j$ on $i$ at each time step (see Fig.~\ref{fig:fldpworks}A).  We refer to such a
system generically as a \emph{forced linear diffusion process} (FLDP).
Intuitively, in such a process the exogenous inputs $z$ diffuse
through the network, transmitted serially by nodes to their neighbors.  To foreshadow, the FLDP leads (under weak conditions) to a single fixed point that describes the long-run behavior of the system, which (when used as a model for social outcomes) we refer to generically as the \emph{forced linear diffusion model} (FLDM).  This is described in detail below.

In the context of social networks, the FLDP has been studied since at
least the mid-20th century.  An early application was the modeling of
status and prestige, which were posited e.g. by \citet{katz:p:1953} to
evolve through an iterative process in which one gains prestige by
associating with or being nominated by prestigious alters, who
themselves gain prestige from their own associations or endorsements.
Assuming a constant, unit ``input'' to the status system leads to the
famous Katz $\alpha$ index.  Power has similarly been posited to
evolve in such a fashion, leading to examples such as the
\citet{salancik:asq:1986} model (in which power inputs are derived
from group memberships) and the \citet{bonacich:ajs:1987} power score
(in which $W$ can be negatively signed, such that individuals become
weaker when their alters are stronger).  Broadly, use of FLDPs in such
applications has focused on deriving equilibrium behavior (as
discussed below), with the equilibrium state of $y$ used as an index
of status, power, or prestige (depending on application).  Such
indices belong to the family of \emph{feedback centralities}, as
summarized e.g. by \citet{koschutzki.et.al:ch:2005}.
A number of properties of these measures, as well as connections with
other centrality measures, were elegantly shown by \citet{friedkin:ajs:1991}

Another important application of the FLDP has been to social
influence.  Most notably, extensive work by Noah Friedkin, Gene
Johnsen, and colleagues has employed FLDP to study the evolution of
attitudes and beliefs in group settings \citep[see
e.g.][]{friedkin.johnsen:jms:1990,friedkin.cook:smr:1991,friedkin:bk:1998,friedkin.et.al:s:2016}.
Beyond theoretical analysis, this work has provided strong empirical
evidence that FLDPs can capture attitude evolution within groups
\citep{friedkin:bk:1998,friedkin.bullo:pnas:2017}.  Friedkin and
Johnsen have further shown conditions on $W$ (corresponding to a
valued graph $G$ of interpersonal influences) that both ensure
convergence of attitudes and that reflect experimentally observed
behavior patterns, and have examined assumptions regarding boundary
conditions (such as the stability of $W$ and $z$), among other
contributions \citep[many of which are summarized in][]{friedkin:bk:1998}.

It should be noted that the FLDP also appears tacitly in other
applications.  For instance, the network autoregressive (NAR) model
\citep{cliff.ord:bk:1973,anselin:bk:1988,doreian:ch:1989} (also known
as the autoregressive spatial autocorrelation model) arises from the
equilibrium of the FLDP, as do the corresponding moving average
models.  Frequently, the origin of this equilibrium condition is not
explained in methodological treatments, and the presumption of an FLDP
giving rise to the observed system may be opaque to practitioners.
(The nature and consequences of these assumptions are discussed
e.g. in \citet{butts:jms:2025} and \citet{friedkin:bk:1998}, Appendix
B.)

\subsection{Equilibrium Behavior and Total Impact}

An attractive feature of the FLDP is the fact that, under relatively
weak conditions, the recurrence of Eq~\ref{eq_lindiff} converges to a
unique fixed point given by
\begin{equation}
y^{(\infty)} = (I-W)^{-1} z, \label{eq_lindiffeq}
\end{equation}
where $I$ is the identity matrix.  A sufficient condition for convergence is that the spectral
radius of $W$, $\rho(W)$, is less than 1.  Intuitively, $W$ is in this
case a long-run contractive mapping, which implies that the diffusion
of $z$ through the network ultimately attenuates; where this does not
hold, there will be at least some choices of $z$ for which $y^{(t)}$
will either oscillate forever or else ``blow up,'' attaining arbitrarily large values on long
timescales.  In most applications of interest, this is unphysical, and
thus $W$ is chosen to ensure convergence.  (See
e.g. \citet{friedkin:bk:1998}, Appendix B, for a discussion in the
context of social influence models.)  One simple strategy - frequently
employed in the context of feedback centrality - is to take
$W=\gamma A$, where $\gamma<1$ is an attenuation parameter.
Convergence is then obtained by setting $\gamma < 1/\rho(A)$.
Substantively, $1/\rho(A)$ can be viewed as the threshold level of
attenuation/interpersonal influence at which the total impact of
individuals on each other through the network will diverge through
runaway amplification; notably, this is a property of the network
structure, and does not depend upon other factors. 

As noted above, most social network applications of the FLDP employ it only via the equilibrium condition of Eq.~\ref{eq_lindiffeq}, assuming (tacitly or otherwise) that timescale separation holds for the evolution of $y$ versus $W$ and that the system of interest is in steady state.  (Indeed, in some cases the FLDP may be invoked only tacitly, or little motivation may be given.)  We thus refer to Eq.~\ref{eq_lindiffeq}, when used as a model for social outcomes, as the forced linear diffusion model (FLDM).  Our interest is particularly in approximating the total impact of one individual upon another within the FLDM, as a function of positional properties and network distance.

The core of the FLDM is the matrix $(I-W)^{-1}$, which we refer to as the \emph{diffusion
  propagator}.\footnote{\citet{friedkin:ajs:1991} works with a slightly rescaled form in the special case of row-stochastic $W$, which he refers to as the \emph{total interpersonal effects matrix.}}  The diffusion propagator can be viewed as a functional of network structure that
acts on the exogenous inputs to the network, $z$, and maps them to the
equilibrium network state.  The substantive interpretation of the
diffusion propagator is aided by noting its series expansion,
\begin{equation}
(I-W)^{-1} = I + W + W^2 + W^3 + \ldots, \label{eq_expansion}
\end{equation}
which converges under the same conditions required for
Eq.~\ref{eq_lindiffeq}.  Recalling that, for an unweighted adjacency
matrix $A$ of graph $G$, the $k$th power of $A$ is an adjacency matrix
whose $i,j$th cell is the number of $k$-walks from $i$ to $j$, we can
see that the sum of Eq.~\ref{eq_expansion} is an adjacency matrix
whose $i,j$th cell contains the total weight of all $i,j$ walks of all
lengths.  For instance, in the simple case for which $w(A)=\gamma A$,
the diffusion propagator has the form
\[
(I-\gamma A)^{-1} = I + \gamma A + \gamma^2 A^2 + \gamma^3 A^3 + \ldots.
\]
The $k+1$th term of the expansion is thus the count matrix
of $k$-walks, downweighted by $\gamma^k$.  Attenuation leads the
impact of any given walk to fall exponentially with length; however,
the number of walks increases rapidly with $k$.  When
$\gamma < 1/\rho(A)$, attenuation eventually dominates, and the total
contributions of all walks to the effect of $j$ on $i$ in equilibrium
remains bounded.

In addition to its value for calculating the equilibrium state
$y^{(\infty)}$, the diffusion propagator is also substantively
interesting for its role in expressing the \emph{total impact} of one
vertex upon another: that is, in equilibrium, \emph{the net
  contribution of node $j$ to the state of node $i$}.  As shown above,
this represents the cumulation of all sequences of interactions
starting at $j$ and ending at $i$, including sequences with cyclic
elements in which individuals reciprocally influence one another.
Total impact is thus a non-local network property, potentially
depending not only on the geodesic connecting $i$ and $j$, but the
broader structure of the interaction network.

Given the form and interpretation of Eq.~\ref{eq_expansion}, it is
natural to wonder whether anything concrete can be said about the
relationship between impact and geodesic distance.  The notion that the
importance of connections falls rapidly with distance is common in
theorizing regarding social networks, with the idea being incorporated
to measures of centrality (e.g., closeness \citep{freeman:sn:1979}), subgroup
definitions (e.g., $k$-cliques \citep{wasserman.faust:bk:1994}), and theories of action
(e.g., network search \citep{lee:bk:1969}).  As seen above, for any fixed degree
of attenuation the impact of any specific walk on the $j,i$ impact is
expected to fall exponentially in distance, which would seem
compatible with the notion that the \emph{total} impact of $j$ on $i$
will also fall when they are at greater remove.  On the other hand,
the increase in the number of walks with distance calls this picture
into question.

The question is consequential: if we wish to understand the effect of
a change in one node's forcing term on others in the network, select
nodes for intervention, or otherwise bound the likely consequence of a
change in network inputs, it is useful to know whether and how total
impact scales with distance.  As we show in the next section, the
intuition that total impact falls with distance is sustained, though
the speed with which this occurs depends on the spectral structure of
the adjacency matrix.  Moreover, total impact is also related to
eigenvector centrality, leading to a useful approximation that
combines properties of $i,j$ distance, centrality, and the global
network structure.

\section{A Simple Model for Total Impact at a Distance} \label{sec:model}

Here, we develop a simple approximation for the total impact of $j$ on
$i$, as a function of the $i,j$ distance in $G$.  For reasons that
will become clear, we will choose to work with a weight matrix
representation $W=w(A)=\gamma B$, where $\gamma<1$.  This is without
loss of generality, since we can simply define $B=W/\gamma$ when not
using e.g. the simple case in which $w(A)=\gamma A$; however, pulling
out an attenuation term $\gamma$ aids in exposition.

Recalling that $W_{ij}$ encodes the immediate impact of $j$ on $i$,
each $i,j$ walk describes a chain of interactions through which $j$ can
affect $i$.  Let the geodesic distance from $i$ to $j$ in $G$ be $d$;
then, obviously, $G$ (and hence $W$) contains no $i,j$ walks of length
$<d$, and it must likewise contain at least one $d$-walk (which must
be a path).  It follows from Eq.~\ref{eq_expansion} that we must have

\begin{align}
  (I-\gamma B)^{-1}_{ij} &= I_{ij} + \gamma B_{ij} + \gamma^2 B_{ij}^2 + \ldots\\ \nonumber
                         &= \gamma^d B^d_{ij} + \gamma^{d+1} B_{ij}^{d+1} + \gamma^{d+2} B_{ij}^{d+2} + \ldots\\ \nonumber
                         &= \gamma^d \left(B^d \left( I + \gamma B  + \gamma^2 B^2 + \ldots\right)\right)_{ij}\\  
                         &= \gamma^d \left(B^d \left(I-\gamma B\right)^{-1}\right)_{ij}.
                         \label{eq_dexpand}
\end{align} 

Now, let $B=S\Lambda S^{-1}$ be the spectral decomposition of $B$,
with $S$ a column-matrix of eigenvectors and $\Lambda$ a diagonal
matrix containing the corresponding eigenvalues.  The corresponding
spectral decomposition of the diffusion propagator is
$(I-W)^{-1}=S Z S^{-1}$, where $Z$ is a diagonal matrix with diagonal
elements $Z_{ii}=1/(1-\gamma\Lambda_{ii})$; this follows from the fact
that the spectral decomposition of $W^k$ is
$W^k=S \gamma^k \Lambda^k S^{-1}$, together with the power series
representation of the diffusion propagator.  Using this, we may
rewrite Eq.~\ref{eq_dexpand} as

\begin{align}
\phantom{(I-\gamma B)^{-1}_{ij}} &= \gamma^d \left(S \Lambda^d S^{-1} S Z S^{-1}\right){ij} \\ \nonumber
                                 &= \gamma^d \left( S \Lambda^d Z S^{-1}\right)_{ij} \\ 
                                 &= \gamma^d \left( S H S^{-1}\right)_{ij}, \label{eq_sexpand}
\end{align}

where $H$ is a diagonal matrix with elements
$H_{ii}=\Lambda_{ii}^d/(1-\gamma \Lambda_{ii})$.

The leading factor of $\gamma^d$ in Eq.~\ref{eq_sexpand} suggests that
total impact with fall exponentially in $d$, but the expression as a
whole is nontrivial.  However, in some cases it may be fairly
straightforward to approximate.  Consider the case for which (1) $B$
has a principal eigenvector $s^{(1)}$ with eigenvalue $\lambda_1$, and
(2) $\gamma$ is close to $1/\lambda_1$.  Then Eq.~\ref{eq_sexpand}
will be dominated by the term associated with $s^{(1)}$, giving us
\begin{equation}
(I-\gamma B)^{-1}_{ij} \approx (\gamma \Lambda_{11})^d/(1-\gamma \Lambda_{11}) s^{(1)}_i {s'}^{(1)}_j, \label{eq_firstorder}
\end{equation}
where ${{s'}^{(1)}}$ is the first row of $S^{-1}$.  

Since $\rho(W)<1$, $\gamma \Lambda_{11}<1$, we see immediately that
Eq.~\ref{eq_firstorder} falls exponentially in network distance.  We
can also see that the strength of this effect depends on the
attenuated principal eigenvalue, $\gamma \Lambda_{11}$, with sharper
declines relative to the $d=1$ baseline when
$\gamma \Lambda_{11} \ll 1$.  By turns, the impact also depends on
the product of $i$'s eigenvalue centrality and a quantity,
${{s'}^{(1)}_j}$, which we may call the \emph{inverse eigenvector
  centrality} of $j$.  Recalling that $W$ reflects incoming impact, we
can understand ${{s}^{(1)}}$ as reflecting the net tendency to
\emph{receive} impact from others, while ${{s'}^{(1)}}$ reflects the
corresponding net tendency to \emph{send} impact.  We would thus tend
to expect the inverse eigenvector centrality to be related to
eigenvector centrality in $W^T$, and indeed, the two are proportional
(where the principal eigenvector exists). In the special case of
symmetric $W$, $S^{-1}=S^T$, and our simple model simplifies to
$(\gamma \Lambda_{11})^d/(1-\gamma \Lambda_{11}) s^{(1)}_i
{s}^{(1)}_j$ ; i.e., in this case the inverse eigenvector centrality
corresponds to the eigenvector centrality.

The above is a \emph{first-order model} for total impact at a
distance.  Note that we can obtain higher-order models by including
additional terms; i.e., for some order $k\le N$,
\begin{equation}
(I-\gamma B)^{-1}_{ij} \approx \sum_{\ell=1}^k (\gamma \Lambda_{\ell\ell})^d/(1-\gamma \Lambda_{\ell\ell}) s^{(\ell)}_i {s'}^{(\ell)}_j. \label{eq_korder}
\end{equation}
This permits corrections for higher-order core-periphery structure
within the interaction.  Care must be used here in the non-symmetric
case, as $S$ and $\Lambda$ will be, in general, complex.  In the
latter case, $S$ and $\Lambda$ will contain eigenvectors and
eigenvalues arising as complex conjugate pairs, and such pairs must be
added together to avoid obtaining a solution with imaginary
components.  Similarly, care must be taken when $W$ has negative
elements, as a principal eigenvector may not exist.  The
approximations of Eqs.~\ref{eq_firstorder},~\ref{eq_korder} remain
valid in this case, but the first eigenvector may not have the
properties usually ascribed to eigenvector centrality.

\section{Assessing the Approximation} \label{sec:assess}

How well does the model of Eq.\ref{eq_firstorder} work?  Here, we
examine this simplified model for impact at a distance, comparing the
approximation with the exact solution for social networks from
educational settings.  We are particularly
concerned with three questions. First, when and to what extent does
the prediction of exponential decline of total impact with network
distance hold? Second, how well does the simple, first-order model do
at approximating total impact? And, finally, are there substantial
gains from using additional, higher-order corrections?  In what
follows, we investigate these questions using the linear form introduced above,
i.e. $W=w(A)=\gamma B$.  This allows us to vary the strength of impact
attenuation, and examine how this affects the behavior of the
influence model.  Note that, because the total impact depends only on
the diffusion propagator - and not the forcing term - our results
depend only on the network (and degree of attenuation), and are
applicable to any model or index that can be expressed in terms of a
FLDM.

\subsection{Data and Methods}

We assess the simplified model as follows.  For each observed network, $A$, in
our study, we construct a series of weight matrices
$W=\gamma/\rho(A) A$, where $\gamma\in (0,1)$ is a decay parameter and
$\rho(A)$ is the spectral radius of $A$.  Thus, $\gamma$ expresses the
strength of immediate impact on a scale that ranges from no impact
($\gamma=0$) to the level of immediate impact at which the feedback
process diverges ($\gamma=1$).  Specifically, we take
$\gamma = 1 - 2^{-k}$, where $k \in 1,\ldots,5$.  To probe both the
symmetric and non-symmetric regimes, we examine all directed networks
separately in raw and symmetrized form (the latter using a weak or
union rule).  In the directed case, each network was oriented so that
$i,j$ ties were interpretively compatible with $j$ having a power or
influence effect on $i$; 
for instance, if $i$ nominates $j$ as a
friend, we assume that $i$ is potentially influenced by
$j$.  

For each prepared network and each $\gamma$ level, we compute the
diffusion propagator (Eq.~\ref{eq_expansion}) from the spectral
decomposition of the weight matrix.  We then compute the first-order
approximation to the total impact via Eq.~\ref{eq_firstorder}.  To
investigate the effect of higher-order corrections, we also compute an
approximation using Eq.~\ref{eq_korder}.  For the symmetric case, or
where two real-valued eigenvectors of maximal modulus eigenvalue are
present, we add the effect of the second-largest modulus eigenvalue to
the approximation.  For directed networks in which no such eigenvector
is present, we take the complex-conjugate eigenvector/eigenvalue pair
whose eigenvalue has the largest modulus, and add both to the
approximation (thus ensuring that the predicted impact remains real).
Finally, we obtained the geodesic distance for each ordered pair of
nodes (treated as infinite where there exists no $i,j$ path).  The
true, first-order approximated, and second-order approximated total
impact scores were analyzed for each dyad or directed dyad (as
applicable).

Data used for this study consists of the following.  We employ the 84
networks from the public use release of the AddHealth data set
\citep{moody:sn:2001}, each containing attributions of friendship from high school
students to other students in their school (including, in some cases,
a junior high school whose graduates typically went on the to the
target high school). These networks vary in size from 25 to 2587
vertices ($\bar{N}=902.4$), with mean degree of 4.38.  As influence is
expected to flow from those viewed as friends to those viewing them as
friends, we leave tie orientations unchanged in our
analyses. We then compute the series of weight matrices for each of these networks. 

All computation was performed using the \texttt{network} \citep{butts:jss:2008a} and \texttt{sna}
\citep{butts:jss:2008b} packages from the statnet \citep{handcock.et.al:jss:2008} library for the R
statistical computing system \citep{rteam:sw:2026}.

\subsection{Results}

Our analysis of the behavior of our approximate model versus the exact
solution to the FLDM reveals the following.

\begin{figure}
  \centering
  \includegraphics[width=0.9\textwidth]{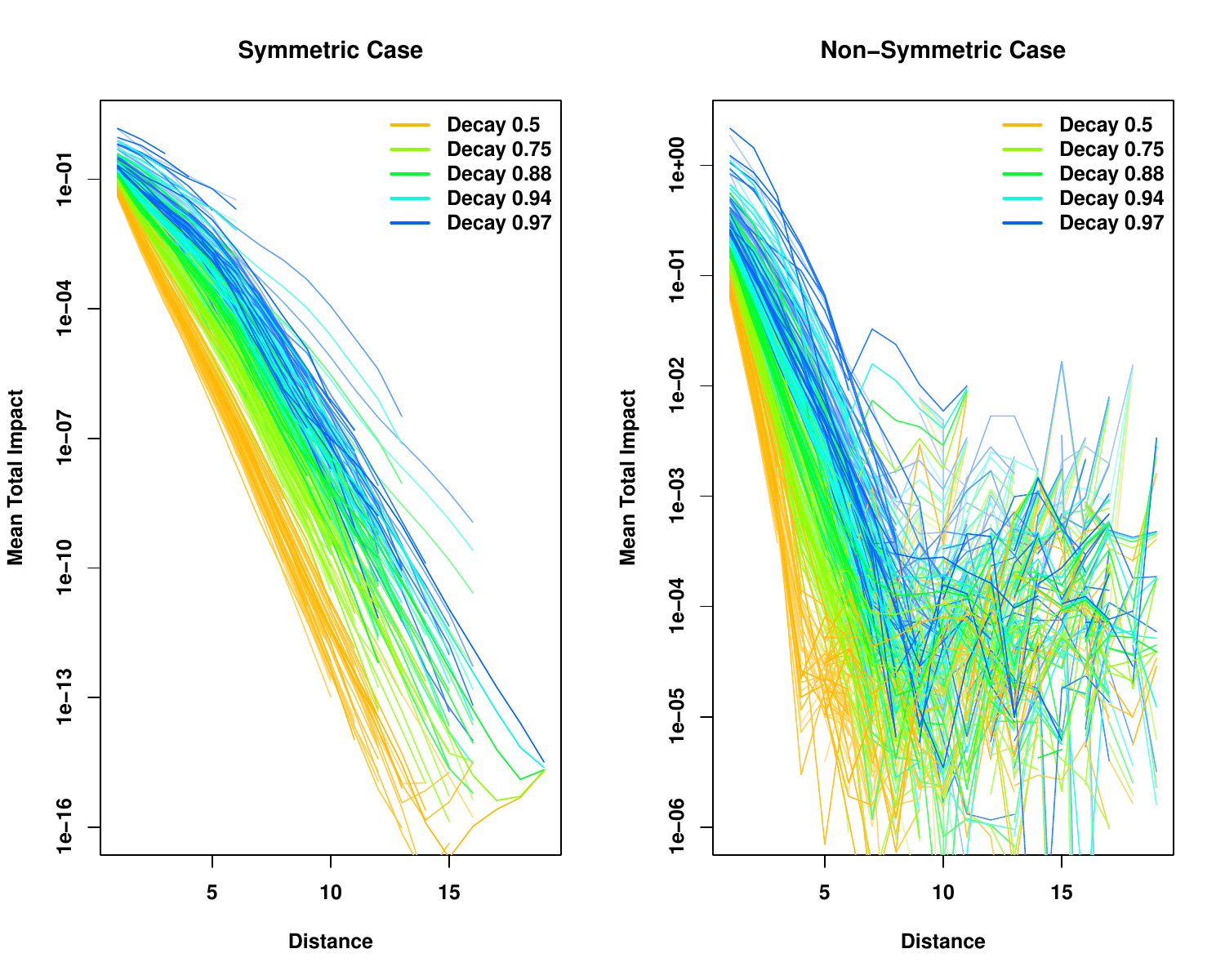}
\caption{Mean total impact as a function of distance from ego, by network.  Left-hand panel shows symmetrized structures for each school, right-hand panel shows directed structures.  Declines are approximately exponential in both cases to distances of 5-10 steps; this trend continues for undirected graphs, while digraphs show disorder at long ranges. }
\label{fig:impbydist}
\end{figure}

\paragraph{Average total impact declines exponentially.} Fig.~\ref{fig:impbydist} shows the mean total impact (as computed using the exact FLDM solution) as a function of distance from ego, for each network at each decay strength.  Our approximation suggests that impact should, \emph{ceteris paribus}, decline exponentially with distance from ego; this condition corresponds to a linear decline on the semi-log plot.  Overall, we find that the exponential approximation works well over the range of primary substantive interest for most network processes, holding to a distance of 5-10 steps in the directed case, and generally to the entire network diameter in the symmetric case; where the pattern does break down (in the directed, long-range regime), the alter-ego impact is usually 1--5 orders of magnitude weaker than the impact of immediate neighbors, implying that these relations contribute little to ego's final state.  Within the exponential regime, we also note that the approximation holds most closely for lower values of $\delta$, with some small deviations at small distances as $\delta \approx 1$.  

It is perhaps less interesting that the exponential approximation fails at long ranges and low impact levels in the directed case, than that the approximation holds so well in the undirected case (with distances as large as 15-20 hops, and impact levels as small as on the order of $10^{-13}-10^{-16}$.  At these ranges, actual impact is substantively negligible, and would in practice be expected to be outweighed by idiosyncratic individual factors.  However, the theoretical impact level is still in exponential decline in this regime.

\paragraph{The first-order model works well for many networks.}

\begin{figure}
  \centering
\includegraphics[page=1, scale=.5]{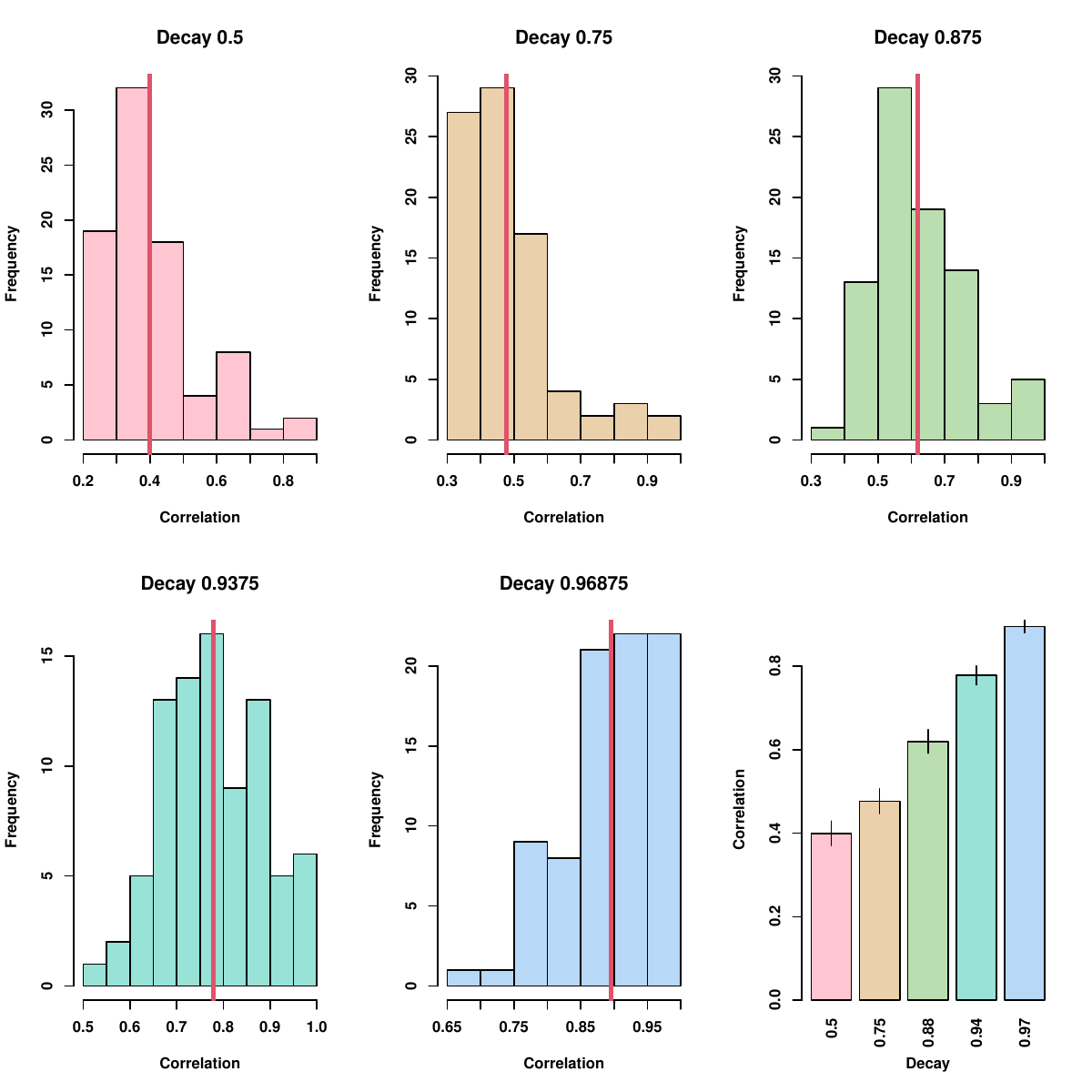}
\caption{Level of correlation between the first order
  approximation and the true geodesic distances for the symmetric
  case for each school. The bottom right figure displays means and standard deviations.}

\label{fig:apxcor_sym}

\end{figure}

\begin{figure}
  \centering
\includegraphics[page=1, scale=.5]{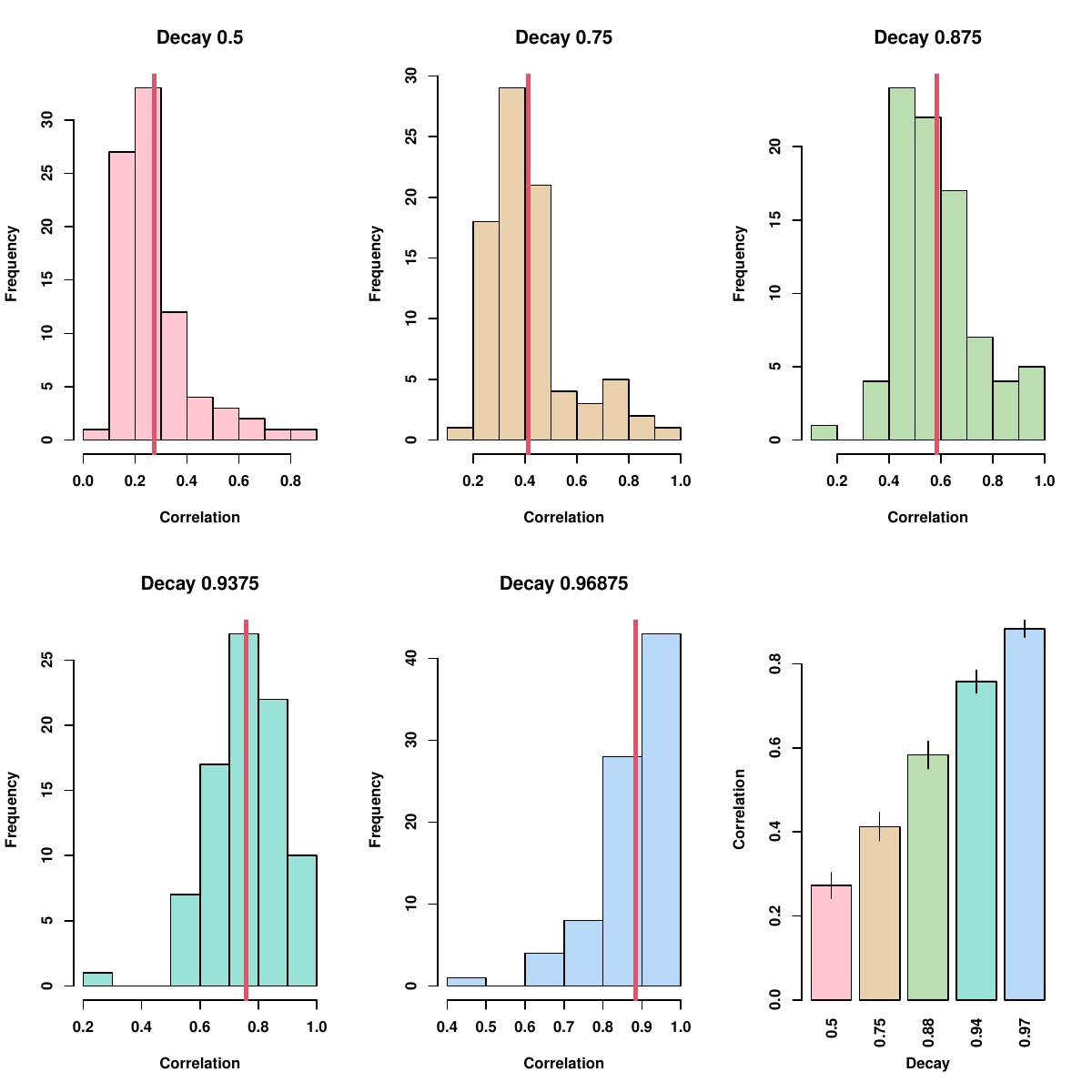}
\caption{Level of correlation between the first order
  approximation and the true geodesic distances for the asymmetric
  case for each school. The bottom right figure displays means and standard deviations.}

\label{fig:apxcorr_asym}

\end{figure}

Overall, the performance of the first order approximation is dependent
on the magnitude of the decay parameter, though many networks show a fairly high correlation even for lower decay values. 

In Figure \ref{fig:apxcor_sym}, we can see the how distribution of
correlation between the true geodesic distances and the 1st order approximations changes as the decay parameter increases for the
symmetric case. As decay increases, the distribution moves from a
right skew (for a decay of .5) to a left skew (when decay is greater
than .9). The center of the correlation distribution also increases,
beginning at a value around .4, and ending at a value just shy of .9.

In Figure \ref{fig:apxcorr_asym}, we can see the how the distribution of
correlation changes as the decay parameter increases for the asymmetric
case. As decay increases, the distribution moves from a right skew
(for a decay of .5) to a left skew (when decay is greater than
.9). This pattern is even more extreme than that observed in the
symmetric case. The center of the correlation distribution also
increases, beginning at a value around .3, and ending at a value just
shy of .9.

For decay parameter values greater than .9, we observe a level of
correlation more than double what occurs when the decay parameter is
less than .5 for both the symmetric and asymmetric cases.

\paragraph{Higher-order improvements help with highly segmented networks.}

\begin{figure}
  \centering
\includegraphics[page=1, scale=.5]{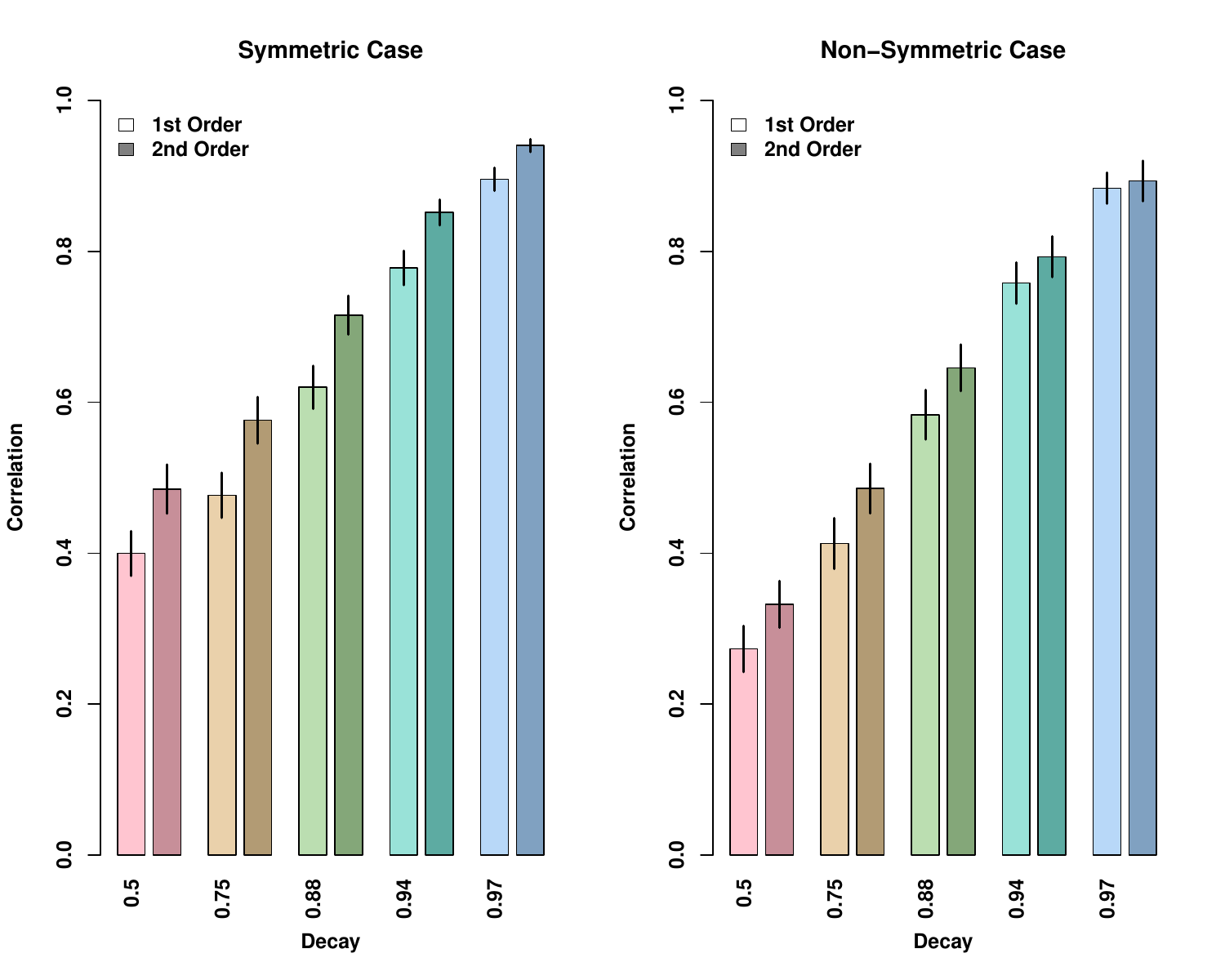}
\caption{Correlation for first and second order eigenvectors as decay
  increases. Note that the degree of correlation is more similar in
  the non-symmetric case than in the symmetric case, particularly as
  decay increases.}
\label{fig:ordercomp}
\end{figure}

In Figure \ref{fig:ordercomp}, we show how the level of correlation
between the true geodesic distances and the first and second
eigenvector approximations behaves as the decay parameter
increases. In both cases the level of correlation increases, with the
non-symmetric case displaying a higher level of similarity between the
two approximations than in the symmetric case.  This reflects the importance of local
segmentation to diffusion: when the decay parameter is low, local structure plays
a greater role, and additional eigenvectors have a larger effect.  By contrast, when
the decay parameter is large - and long paths have a substantial contribution - the
global structure arising from the principal eigenvector account for more of the 
total impact.

\section{Discussion} \label{sec:discuss}

\subsection{Generality of the FLDM}

As noted above, \citet{wang.et.al:s:2026} report exponential decay of total impact from network interventions (measured relative to network distance at the time of intervention) in a more elaborate model of social influence incorporating both edge selection and alternative influences on behavior.  Here, we have shown that the same pattern is expected to arise under linear diffusion, under weak conditions.  Is the correspondence accidental?  We suggest otherwise.  Although the exact processes governing behavior change in a stochastic actor-oriented model (as in Wang et al.) or similar agent-based models are much more elaborate than in the linear diffusion model, we can nevertheless view the linear model as a first-order approximation to these more complex cases.  So long as rate of network change in such a model is relatively slow compared to behavior change, the diffusion process will be well-approximated by a perturbation to the static model \citep{butts:jms:2025}, and we would thus expect many of the properties governing the latter to approximately hold, as well.  To the extent that many real systems fall into this regime, the FLDM may have substantially greater applicability than its simple structure might suggest (at least for providing qualitative insight).  That said, a systematic comparison of how strongly empirically calibrated models of interpersonal influence deviate from the linear model remains an open problem, as does the robustness of the FLDM to violation of many of its core assumptions.  This would seem to be a fruitful direction for further work.

\subsection{Implications for Network Interventions}

Workers in many settings (notably public health) have proposed interventions that target
specific individuals for intervention. Interventions targeting individuals  have been applied broadly across various behavioral domains
from adolescent substance use \citep{lakon2025simulating, wang2022insight} to community engaged approaches
\citep{ayala2011effects}.  These interventions are often delivered within
bounded social systems such as schools for children and adolescents,
students in universities, people in community organizations,
coalitions and even the broader contexts of communities such as
Project Northland \citep{perry2002project}. Such interventions have
been studied within numerous populations \citep{eng1993save, valente2012network, lakon2025simulating}.

The overall goal of such interventions is to
modify the behavior of not just the specific individuals, but many
others members of a network via the influence of the targeted on the untargeted. One intuition for targeting specific
individuals is that some members of a network are more likely to have
additional effects on their peers via some influence process.

As only a limited number of individuals can be practically targeted due
budgetary or other constraints, a number of challenges
emerge. Subject to relevant constraints, an ideal intervention should
impact a large proportion of the population of interest, both directly
(the individuals intervened on), and indirectly (via the influence
process).

Despite being a simplified model of influence, the FLDM approximation 
provides useful insights which help address this practical issue for
planners of interventions. 

Long-range effects from interventions, where two individuals in a
network are distant from one another, are predicted to be small, due to the
exponential decline of influence with increasing distance. An
intervention should therefore focus on those nodes near a targeted
ego, as spillover effects should largely dissipate after the
second-order neighbors of said target.

Influence in a transposed\footnote{Transposed, as in a friendship
  network, $i$ nominating $j$ as a friend suggest that $j$ influences
  $i$.} influence network should be strongest for nodes with high
eigenvector centrality. They should also have the greatest influence
on other individuals with a high eigenvector centrality. It then
follows that indirect effects will primarily impact the core(s) of a
network, rather than the periphery.

The greater the direct influence of a particular node on another, the
stronger and more predictable the indirect effects will also be. This
suggests that there are substantial benefits to approaching the
highest level of influence, though in practice this is unfortunately
unlikely to be easily controlled.

These three implications from the FLDM provide potentially useful guidance
for practitioners when designing intervention studies, and should
increase the accuracy of the predicted effectiveness of an
intervention where using more elaborate models is infeasible due e.g. to data constraints.

The FLDM approximation approach can give insight into how the social impact from a
specific salient individual in a network can affect those along a
chain of alters. As such, it can be potentially used to inform existing network
intervention strategies targeting specific individuals.  For instance, this
approach can provide insight into understanding the possible impact of interventions such as
 peer-based support
network interventions for smoking cessation \citep{lakon2016mapping},
creating buddy systems for smoking cessation \citep{may2000social}, and
in leveraging peoples’ existing social network ties to build new ties
such as in the seminal Tenderloin Senior Outreach Project (TSOP) which
created new social structures and resources for impoverished and
socially isolated seniors in the Tenderloin area of San Francisco
\citep{minkler1986building}.  Considering how perturbations to an existing network might affect interpersonal impact can help guide intervention strategies
that involve developing new social ties to increase network size and the
opportunities for more collective resources, which was a critical
intervention strategy in the Tenderloin Senior outreach Projects
\citep{minkler1986building}.

Insights from the linear model can also inform the strategy of identifying and
diffusing interventions through salient community members who operate
as natural helpers or leaders in communities, such as the lay health
advisor approach \citep{eng1993save} and the promotor(a) model
\citep{ayala2011effects}. Such people often occupy a central node or
bridging tie position in their networks, which are social positions
which confer advantage and ready access to a diversity of resources
and the ability to influence many others. As such, these distinguished
members of a community are in a position to greatly catalyze behavior
change at the individual level with potential for scale up to the
community level.  As we observe here, those expected to have, on average
maximum interpersonal impact are those with maximum inverse eigenvector
centality, and those most readily influenced by them are those nearby
with maximum eigenvector centrality.  While the notion of attempting 
to maximize secondary influence by targeting individuals with high
eigenvector centrality in the influence-spreading network (as opposed 
to its influence-receiving transpose) is well-established, the FLDM
provides not only a clear theoretical rationale for the importance
of eigenvector centrality per se, but also suggests how \emph{influencable}
individuals may be found, and how close they must be to receive 
substantial secondary influence from an intervention.  These insights
complement those attainable through other methods (e.g., detailed
simulation studies).

\section{Conclusion} \label{sec:conclusion}

We have demonstrated that under a forced linear diffusion
model, it is possible to approximate both direct and indirect impact
between nodes in a network, and used this result to estimate the total
impact one node has on another. We found that the degree of indirect
influence decreases exponentially as the distance between two nodes
increases.  More significantly, this result follows from a simple
analytical model, rather than a simulation study.

Some practical implications of this model include the importance of
prioritizing nodes with high eigenvector centrality, as well as the
immediate near-neighbors of selected targets, as these nodes are most
likely to be impacted by an intervention on the selected targets. This
model also provides computational advantages, as the traditional
approach of computing $(I-W)^{-1}$ (Eq~\ref{eq_lindiff}) is costly for
large networks, while our method only requires that geodesic distance
and the principal eigenvector be calculated.

The ability to more effectively select targets for an intervention
should greatly increase both the ability to discover interventions
whose impact via indirect influence was concealed by low power
studies, as well as enable beneficial interventions at a lower cost --
an important feature in these austere times.

\bibliography{bibhome/ctb}


\end{document}